\documentclass[12pt]{article}
\usepackage{epsf}
\newcommand{\ba}{\begin{array}}
\newcommand{\ea}{\end{array}}

\newcommand{\beq}{\begin{equation}}
\newcommand{\eeq}{\end{equation}}
\newcommand{\bea}{\begin{eqnarray}}
\newcommand{\eea}{\end{eqnarray}}
\newcommand{\beal}{\setcounter{letter}{1} \begin{eqnarray}}
\newcommand{\eeal}{\addtocounter{equation}{1} \end{eqnarray}}
\newcommand{\none}{\nonumber \\}
\newcommand{\req}[1]{Eq.(\ref{#1})}

\newcommand{\larrow}{\,\,\,\,\hbox to 30pt{\rightarrowfill}
\,\,\,\,}
\newcommand{\slarrow}{\,\,\,\hbox to 20pt{\rightarrowfill}
\,\,\,}

\begin{document}

\begin{titlepage}
\renewcommand{\thefootnote}{\fnsymbol{footnote}}
\renewcommand{\baselinestretch}{1.3}

\begin{center}
{\large {\bf Approach to Equilibrium in the
Micromaser}}
 \\ \medskip  {}
\medskip

\renewcommand{\baselinestretch}{1}
{\bf D. Leary${}^{1,2}$, S. Yau${}^1$, M. Carrington${}^{3,4}$,
  R. Kobes${}^{1,4}$\\
 and G. Kunstatter${}^{1,4}$
\\}
\vspace*{0.50cm}
{\sl
${}^1$ Dept. of Physics\\
 University of Winnipeg\\
Winnipeg, Manitoba, Canada R3B 2E9\\ }
\vspace*{0.50cm}
{\sl
${}^2$ Current Address:  Dept. of Physics, \\
Memorial University\\
St. John's, Newfoundland Canada\\ }

\vspace*{0.50cm}
{\sl
${}^3$ Dept. of Physics\\
 Brandon University\\
Brandon, Manitoba, Canada R7A 6A9 \\ }
\vspace*{0.50cm}
{\sl
${}^4$ Winnipeg Institute For Theoretical Physics\\
Winnipeg, Manitoba, Canada R3B 2E9\\ }
\vspace*{0.50cm}
\end{center}

\renewcommand{\baselinestretch}{1}

\begin{center}
{\bf Abstract}
\end{center}
{\small
We examine the approach to equilibrium of the micromaser.
Analytic methods are first used to
show that for large times (i.e. many atoms) the convergence is governed
by the next to leading eigenvalue of the corresponding discrete
evolution matrix. The model is then studied
numerically.
The numerical results confirm the phase structure
expected from  analytic approximation methods and  agree for
large times with the analysis of Elmfors {\it et al} in terms of the
``continuous master equation''. For short times, however, we  see
evidence for interesting new structure not previously reported
in the literature.
}
\vfill
\hfill \today \\
\end{titlepage}

\section{Introduction}

The micromaser\cite{maser} provides an excellent theoretical and experimental
testing ground for many fundamental properties of cavity quantum
electrodynamics and quantum mechanics in general. The physical situation under
consideration
consists of a superconducting, high $Q$ cavity, that is being traversed by
a low intensity beam of two state atoms. The atoms interact with the
electromagnetic field in the cavity via an electric dipole interaction.
 The
dynamics of the atom cavity system is  well described by the Jaynes-Cummings
model\cite{jc}. If the cavity transit time $\tau$ is short compared to the
average time
$T$ between
atoms, there is effectively only one atom in the cavity at a time and the atoms
in the beam
interact with each other only via their residual effect on the electromagnetic
field. For example if the atoms enter the cavity preferentially in their
excited state, and emit a photon, the photons tend to build up inside the
cavity, and each successive
 atom sees a stronger photon field when it enters the cavity.
This ``pumping'' is responsible for the evolution of the system into a
microwave laser, or ``maser''.

Three independent time scales determine the overall dynamical
behaviour: the time interval $T$ between consecutive atoms, $\tau$ the
time spent by each atom inside the cavity and $1/\gamma$, the
characteristic photon decay time $\gamma$ inside the cavity.
   An important physical
quantity
is the dimensionless ``pumping rate'' $N=R/\gamma$, where $R=1/T$ and
$\gamma$ is the characteristic photon decay time of the cavity.  $N$
can be thought of as the number of atoms that pass through the cavity
in a single photon decay time.  When both damping and
pumping are present, the photon distribution inside the cavity asymptotically
approaches
a steady state distribution.  The
details of the steady state (equilibrium) distribution depend on the time
 $\tau$ that the atom spends in the cavity as well as the dimensionless
pumping rate $N$.

Although much  work has been done on the equilibrium properties of
 this system,
 to the best of our knowledge there
has not been a systematic analysis of the initial stages
of the approach to equilibrium, which
in principle can be important in determining the outcome of very low flux
experiments.  The purpose of the present work is to examine numerically
this approach to equilibrium. In particular, we will see how varying the
physical parameters affects the rate at which equilibrium is reached: i.e. how
many atoms must pass through the cavity before a steady state photon
distribution is established. In a recent paper, Elmfors {\it et al}\cite{Pers}
looked
at long time correlations in the outgoing atomic beam and their relation to
the various phases of the micromaser system. The properties they
considered were  associated with the equilibrium configuration
of the cavity photon distribution, but
there is a close connection between
the correlation functions considered in \cite{Pers}  and the near equilibrium
dynamical behaviour
that we will be examining.  As we will show, our results agree
with those of \cite{Pers} in the appropriate (i.e large $N$) limits.

The paper is organized as follows: In Section 2 we review the JC model and
its application to the physical situation at hand. In particular we
derive transition matrix $S$ that governs the master equation for the dynamical
evolution of the
photon distribution inside the cavity. We also derive the expression for the
probability $P(+)$ of finding the atom in the excited state.
 In Section 3 we show that the
approach to equilibrium of the photon distribution and of the physically
measured $P(+)$ is governed by the
leading eigenvalues of $S$. We compare
our  theoretical analysis to that of Elmfors {\it et al}, who looked at
correlation functions
instead of the photon distribution directly. In Section 4, we describe the
numerical experiment that we use to analyze the approach to equilibrium,
and compare our results to our theoretical analysis and to that of Elmfors
{\it et al}.  Section 5 closes with a summary and conclusions.

\section{The Jaynes-Cummings Model}

We consider atoms with two possible states $|\pm\rangle$ with energy difference
\beq
E_+-E_- = \hbar \omega_a
\eeq
For a high $Q$ cavity,  the electromagnetic field
is well approximated by a single mode, with energy $E_c=\hbar \omega_c$.
For simplicity we assume that the cavity is tuned so that its fundamental
frequency is equal to that of the atom:
\beq
\omega_a=\omega_c =\omega
\eeq
For a single atom traversing the cavity, the dynamics of the atom-cavity system
is governed
 by the JC
Hamiltonian.
\beq
H = \omega a^\dagger a + {1\over 2} \omega \sigma_z + g(a \sigma_+ + a^\dagger
\sigma_-)
\eeq
where $g$ is the coupling constant, $a^\dagger$ ($a$) are the photon
creation (annihilation) operators and
$\sigma_ \pm = {(\sigma_x \pm i\sigma_y) \over 2}$ are operators which
raise and lower the atomic states
($\sigma_x$, $\sigma_y$, and $\sigma_z$ are the Pauli matrices). In the
absence of the dipole interaction (i.e. when $g=0$) the atom-plus-field
energy eigenstates are  $|n,s\rangle$, where $n=0,1,...$ is
the  the photon number  and $s=\pm$ for the two atomic levels.  When $g$ is
non-zero the system makes transitions between the energy eigenstates of
the non-interacting system with probabilities,
\bea
| \langle n,-| e^{ -iH t} |n,- \rangle |^2  &=& 1 - q_n(\tau)\nonumber \\
| \langle n-1,+| e^{ -iH t} |n,- \rangle |^2  &=& q_n(\tau)\label{q}\\
| \langle n,+| e^{ -iH t} |n,+ \rangle |^2  &=& 1 - q_{n+1}(\tau)\nonumber\\
| \langle n+1,-| e^{ -iH t} |n,+ \rangle |^2  &=&  q_{n+1}(\tau)\nonumber
\eea
These probabilities are expressed in terms of the quantity,
\beq
q_n(\tau) = \sin^2 \left(
g \sqrt{ n}\tau \right)
\eeq
  This
is a completely solvable quantum mechanical system. We suppose that the atom/
cavity states are uncorrelated at $t=0$, so that
\bea
|\psi\rangle &=& |\psi_{atom}\rangle \otimes |\psi_{cav}\rangle\nonumber\\
 &=& (\alpha|+\rangle+\beta|-\rangle)\otimes (\sum_n C_n|n\rangle)
\eea
The interaction between the  atom
and electromagnetic field causes the states to be entangled. The exact result
for the wave function after an interaction time $t$ is:
\bea
|\psi(t)\rangle &=& \sum_n\left[\left(\alpha C_n \cos(g\sqrt{n+1}t) -i\beta
C_{n+1}
  \sin(g\sqrt{n+1}t) \right)\right.|n,+\rangle\none
& &\qquad + \left.\left(-i\alpha C_{n-1} \sin(g\sqrt{n}t) +
   \beta C_n \cos(g\sqrt{n}t)\right)|n,-\rangle\right]
\label{eq: exact wave function}
\eea
We now define $P_{n,s}(t)$ as the probability of finding the atom
in the state $s$, and $n$ photons in the cavity. Specifically, one has:
\bea
P_{n,+}(t) &=&\left\langle n,+|\psi(t)\rangle\right|^2\none
   &=& aP_n (1-q_{n+1}(t)) + bP_{n+1}q_{n+1}(t)\none
   &=&   [a(1-q_{n+1})\delta_{n,m} + bq_{n+1}\delta_{m,n+1}]P_m \nonumber \\
&=:& M(t,+)_{nm} p_m \label{Mp}\\
P_{n,-}(t) &=&\left|\langle n,-|\psi(t)\rangle\right|^2\none
   &=& aP_{n-1}q_{n}(t) + bP_{n}(1-q_{n}(t))\none
&=& [aq_n \delta_{m,n-1} +b(1-q_n) \delta_{n,m}]P_m \nonumber \\
&=:& M(t,-)_{nm} P_m \label{Mm}
\eea
where $a=\alpha^*\alpha$ ($b=\beta^*\beta$) is the probability that the atom
entered the cavity
in the excited (lower) state, while $P_n=C_n^*C_n$ is the probability
that there were $n$ photons in the cavity initially.
It follows directly that the probabilities ${\cal P}_+(t)$ and
${\cal P}_-(t)$  of finding the atom
in the upper and lower states, respectively, for unknown cavity state, are:
\bea
{\cal P}_+(t)&=&\sum_n\left|\langle n,+|\psi(t)\rangle\right|^2\none
 &=& \sum_n\left(a P_{n}
(1- q_{n+1}(t)) + b P_{n+1}q_{n+1}(t)\right)
\label{eq: P+}\\
{\cal P}_-(t)&=&\sum_n\left|\langle n,-|\psi(t)\rangle\right|^2\none
  &=& \sum_n\left(a P_{n-1}
q_n(t) + b P_{n}(1-q_n(t))\right)
\label{eq: P-}
\eea
Eqs.(\ref{eq: P-}) and (\ref{eq: P+}) can be written in matrix form:
\bea
{\cal P}_s(t) = \sum_{n,m}M(t,s)_{nm} P_m \label{calP2}
\eea
where $M(t,s)_{nm}$ is defined in Eqs(\req{Mp}) and (\req{Mm}) above.

 Conversely, if we
are not interested in determining the state of the atom then
the probability of finding exactly $n$ photons in the cavity is\footnote{We
henceforth adopt the convention that repeated indices are to be summed, unless
stated otherwise.}:
\bea
P_n(t)&=&  P_{n,+}(t) + P_{n,-}(t)\none
&=& M(t,+)_{nm} P_m + M(t,-)_{nm} P_m \nonumber \\
&\equiv&  M(t)_{nm} P_m
\label{eq: photon distn}
\eea
where
\bea
M_{nm} = aq_n\delta_{n,m+1} + b q_{n+1}\delta_{n+1,m} + (1 - aq_{n+1} - b
q_n)\delta_{n,m} \label{M}
\eea
Eqs.(\ref{eq: photon distn}) and (\ref{M}) give the master equation for
the time evolution of the photon distribution in the presence of the
atom-cavity interaction, without thermal dissipation.
The first term in the transition matrix $M$ gives the probability that an
$n$-photon state occurs through decay of the excited atomic state in
interaction with a cavity containing $n-1$ photons.  It is given by the product
of $a$ (the probability for the atom to be in the excited state) times
$p_{n-1}$ (the probability that the cavity contains $n-1$ photons) times
$q_n(\tau)$ (probability for a transition between an unperturbed eigenstate
$|n-1,+\rangle$ and an unperturbed state $|n,-\rangle$.  Similarly, the second
term comes from the excitation of the atomic ground state, and the third term
is the contribution from processes that leave the atom unchanged.

The above analysis assumes that the  system under consideration is
in a pure quantum state.
In a realistic experiment both the atom and cavity would be described by
a density matrix representing a mixed state. We will now show that under
some simple assumptions the above formulas also apply to the more realistic
case.
Let $\hat{\rho}(t)_{aC}$ denote the density matrix
describing the atom/cavity system at time $t$.
Operator expectation values are given by,
\bea
\langle \hat{{\cal O}}\rangle = {\rm Tr}_{aC}(\hat{\rho}(t)_{aC} \hat{{\cal
O}})
\eea
where the subscript $aC$ indicates that the trace is over both the atomic and
the cavity states. If we restrict ourselves to the measurement of observables
that involve only atomic operators, the expectation value of such an operator
is given by,
\bea
\langle \hat{{\cal O}}(t)\rangle = {\rm Tr}_a (\hat{{\cal O}}(t) {\rm Tr}_C
\hat{\rho}(t)_{aC}) \equiv {\rm Tr}_a (\hat{{\cal O}}(t) \hat{\rho}(t)_{red})
\eea
where the operator
\bea \hat{\rho}(t)_{red} = {\rm Tr}_C\hat{\rho}(t)\label{rhored}
\eea
is the trace over the cavity states of the total density matrix and is called
the reduced density matrix of the system.  To determine the expectation values
of atomic operators at arbitrary times we need to know $\rho(t)_{red}$ for any
time $t$.

Our system consists of a series of atoms that enter a cavity containing
electromagnetic radiation.  We require that the time $T$ between atoms, and the
photon decay time $1/\gamma$, are much larger than the time that any given atom
spends in the cavity ($T\gg \tau$, $1/\gamma \gg \tau$); equivalently
we assume that the time scale of interactions within the reservoir is much
smaller than the time scale over which we want to consider the evolution of the
system.  Under this condition the density matrix  for the initial state of the
system can be factored into a product of density matrices for the cavity and
the individual atoms:
\bea
\hat{\rho} = \hat{\rho}_C \otimes \hat{\rho}_{a}
\eea
Substituting into \req{rhored} we have,
\bea
\hat{\rho}(t)_{red} = {\rm Tr}_C (\hat{\rho}(t)_a \otimes \hat{\rho}(t)_C)
\eea
We treat the cavity as a reservoir and sum over the large number of reservoir
states to obtain the ensemble averaged density matrix,
\bea
\hat{\rho}_C \rightarrow \hat{\bar{\rho}}_C = {\rm lim}_{{\cal N}\rightarrow
\infty} \frac{1}{{\cal N}} \sum_{n=1}^{{\cal N}}\hat{\rho}_n = \sum_n^{\infty}
p_n |n\rangle \langle n| \label{rhoC}
\eea
where $p_n \ge 0$ and $\Sigma_{n=0}^{\infty}\,\,p_n = 1$.
We assume that the incoming atoms are uncorrelated and have initial states that
can be represented as a diagonal mixture of excited and unexcited states with a
density matrix of the form,
\bea
\hat{\rho}_{a}= \left(\begin{array}{cc} a & 0 \\ 0 & b \\ \end{array} \right) =
a|+\rangle \langle +|\,\,+\,\, b|-\rangle \langle -| \label{rhoa}
\eea
where $a,b \ge 0$ and $a+b=1$.  Physically, $a$ is the probability that the
atom is  initially in the excited state, and $b$ is the probability that the
atom is initially in the ground state.  Combining \req{rhoC} and \req{rhoa} we
have
an expression for the atom-cavity density matrix at the time the
atom first enters the cavity:
\bea
\hat{\rho}(0) &=& \sum_n\left( ap_n |n+\rangle \langle n+| \,\,\,+\,\,\, bp_n
|n-\rangle \langle n-|\right)
 \label{rhored0}
\eea
To study the time evolution of the atomic variables, we need the time
dependent reduced density matrix at time $t$.
A straightforward calculation reveals that
\bea \hat{\rho}(t)_{red} &\equiv& Tr_C\left(e^{-iHt}\hat{\rho}(0)
e^{iHt}\right)
\none
 &=& \left( \begin{array}{cc} {\cal P}_+(t) & 0 \\ 0 & {\cal P}_-(t) \\
\end{array} \right)
 \label{rhored2}\eea
with
${\cal P}_s(t)$ given in \req{eq: P+} and \req{eq: P-}.
Similarly, if we are interested in measuring only cavity observables, we
must consider the reduced density matrix for the cavity, which after time, $t$
is given by:
\bea
\rho_{C,red}(t) &= & \hbox{Tr}_a\left(e^{-iHt}\hat{\rho}(0) e^{iHt}\right)\none
 &=& \sum_n  P_n(t)|n\rangle\langle n|
\label{eq: reduced cavity}
\eea
where $P_n(t)$ is given by \req{eq: photon distn}. Thus if the initial
atomic and cavity density matrices are diagonal, then they both remain
diagonal, and they give rise to precisely the same master equation for the
photon  probability distribution as in the case of pure states.

\req{M} can be modified to include the effects of thermal
 damping. Suppose the photon distribution inside the cavity initially is
$p^{(0)}$. An atom enters the cavity and exits after an interaction
 time $\tau$. Assuming that  $\gamma\tau\ll1$, we neglect
damping during the atom-photon interaction.  The probability distribution for
the photons
just before the next atom enters the cavity is\cite{Pers}:
\bea
p(T) = e^{-\gamma L_C T}M(\tau)p^{(0)}
\eea
where $T$ is the time between atoms and
\bea
(L_C)_{nm} = (n_b+1)(n\delta_{n,m} - (n+1)\delta_{n+1,m}) +
n_b((n+1)\delta_{n,m} - n\delta_{n-1,m})\label{LC}
\eea
We can also take into account the fact that atoms in the beam arrive at time
intervals that are Poisson distributed, with an average time interval of
$T=1/R$ between them.  Multiplying by the distribution function ${\rm
exp}(-RT)\,\,RT$ and integrating we find the averaged photon distribution just
prior to the
arrival of the second atom to be:
\bea
\langle p(T)\rangle_T &=& Sp^{(0)}\label{master} \\
S &=& \frac{1}{1+L_C/N}M \label{S}
\eea
This is the form  of the master equation that we will use to describe the
dynamics of the photon distribution inside the cavity. We will refer
 to it as the discrete master equation.

The analysis in \cite{Pers} starts from a different master
equation, which is called the continuous master equation. To derive
the continuous master equation, Elmfors {\it et al} consider a situation where
the flux of incoming atoms is large enough that the atoms have Poisson
distributed arrival times, so
that each atom has the same probability $Rd\,t$ of arriving in an infinitesimal
time $d\,t$. They  further assume
 that the interaction within the cavity takes place in a time much less than
this time interval ($\tau\ll d\,t$) which means that the interaction is
essentially instantaneous.  The contributions from damping and pumping during
the time
interval $d\,t$ can  then be considered separately.  The contribution from the
damping is exactly the same as before [\req{LC}].  The contribution from
pumping has the form,
\bea
(dp)_{pump} = (M-1)R p d\,t
\eea
 where $M$ is given in \req{M} as before.   The
continuous form of the master equation is obtained by combining the
two contributions,
\bea
d\,p &=& \{-\gamma L_{C} p + (M-1)R p\} d\,t =: -\gamma L p d\,t
\label{cont.master}\\
L &=& \left( n_{b}+1 \right)
\left( n\delta_{n,m} -\left( n+1 \right) \delta_{n+1,m} \right)
+ n_{b}\left( \left( n+1 \right) \delta_{n,m} - n\delta_{n,m+1}
\right)\nonumber \\
&+&
 N \left( \left( aq_{n+1}+bq_{n} \right) \delta_{n,m} - aq_{n} \delta_{n,m+1}
- bq_{n+1} \delta_{n+1,m} \right)\label{L}
\eea
It is expected that the continuous and discrete master equations agree
when the number of atoms per photon decay time is very large.\footnote{Note
that the atomic flux must still be small enough so that effectively only one
atom is in the cavity at a time.}. In particular,  when the
thermalization time scale $1/\gamma$ is much greater than the time $T$ between
atoms, the large time (many atom) dynamics should agree. This correspondence
will be
verified explicitly below.

\section{ Eigenvalue Problem and Approach to Equilibrium}

Recall the physical system we are considering.  We inject a series of atoms
into a cavity.  The time, $T$, between the atoms is much greater than the time
an individual atom takes to pass through the cavity $\tau$ so that there is
never more than one atom in the cavity at one time.  Each atom
 interacts with the cavity's electromagnetic field as it passes through the
cavity, and consequently, the photon distribution changes repeatedly. According
to the discrete master equation \req{master} after $k$ atoms have passed
through the cavity, the photon distribution, $p{(k)}$ is:
\beq
p{(k)}= S^k p{(0)},
\label{master2}
\eeq
where $p{(0)}$ is the initial photon distribution. Equilibrium occurs when the
photon distribution is no longer changed by the transition matrix $S$. That is
\beq
Sp^{eq}= p^{eq}
\label{equilibrium}
\eeq
and the equilibrium photon distribution $p^{eq}$ is an eigenvector of $S$ with
eigenvalue 1. We expect that as $k\to \infty$, $p{(k)}\to p^{eq}$, and it
is precisely this approach to equilibrium that we wish to investigate.

As discussed in \cite{Pers} the approach to equilibrium is governed by the
eigenvalues of $S$. Before proving this, we need some preliminaries.
 The right eigenvectors of the matrix $S$ [\req{S}] are written $p^{(l)}$, the
left eigenvectors are $u^{T(l)}$ and the eigenvalues are $\kappa^{(l)}$,
\bea
S p^{(l)} &=& \kappa^{(l)} p^{(l)}\label{eigenvector1} \\
u^{T(l)} S &=& u^{T(l)} \kappa^{(l)}\label{eigenvector2}
\eea
\req{equilibrium} implies that there is an eigenvector with eigenvalue
unity. As we will
see,
all other eigenvalues must be less than one in order for the system to
be stable.  For convenience, we label the eigenvalues by size;
\beq
\kappa^{(1)}=1
 > \kappa ^{(2)} > \kappa^{(3)}. . . .
\eeq
With this labelling, $p^{(1)}\equiv p^{eq}$ and
\bea
Sp^{(1)} &=& p^{(1)}
\label{right equil}\\
u^{T(1)} S &=& u^{T(1)}
\label{left equil}
\eea
{}From \req{M}, \req{LC} and \req{S} it follows that $u^{T(1)}$ is a
vector with all components equal to one. To see this, note that the
transformation matrix $S$ must preserve the norm
of the probability distribution. It therefore follows that
\beq
\sum_{n,m} S_{nm}p_m = \sum_n p_n
\eeq
for all $p_n$, which in turn requires:
\beq
\sum_n S_{nm} = 1
\eeq
for all $m$. This
is equivalent  to \req{left equil} with
$u^{T(1)}_n=1$.

We will use the fact that the  similarity transform of the form
\bea
T_{ml} = p_m^{(l)}\label{p}
\eea
 diagonalizes the matrix $S$ and that
 the inverse of $T$ is given by the left eigenvector:
\beq
T_{lm}^{-1} = u_m^{T(l)}
\label{u}
\eeq
 Finally, from \req{p} and \req{u} we have,
\bea
u_m^{T(l)} p_m^{(a)} = T_{lm}^{-1} T_{ma} = \delta_{la}
\eea
These results allow us to write, the evolution matrix as,
\bea
S_{nm} = \Sigma_{l=1}^{\infty} \kappa^{(l)} p_n^{(l)} u_m^{T(l)}
\eea  It is easy to verify that this expression satisfies the eigenvector
equations \req{eigenvector1} and \req{eigenvector2}.

In order to investigate the approach to equilibrium we start with the fact that
\beq
p(k+1) = S p(k)
\eeq
Now define
\beq
dp(k) = p(k+1)-p(k) = (S - 1) p(k)dk
\label{discrete evolution}
\eeq
where $dk = (k+1)-k=1$.
This is the discrete  version of the continuous evolution
equation \req{cont.master}. In the limit of large $k$, it can be treated
as a differential equation. In order to integrate it, we first define
\beq
Q(k) = T^{-1}p(k)
\eeq
so that \req{discrete evolution} reads in component form:
\beq
dQ_n(k)= (\kappa^{(n)}-1) Q_n(k)dk
\label{Q}
\eeq
 \req{Q} can be trivially integrated to give
\beq
Q_n(k)= Q_n(0) e^{-(1-\kappa^{(n)})k}
\label{Q soln}
\eeq
The solution for the photon distribution after the $k$th atom is therefore:
\bea
p_n(k)&=& \sum_{m,l}T_{nm}\exp(-(1-\kappa^{(m)})k)T^{-1}_{ml}p_l(0)\none
  &=& \sum_{m,l} p^{m}_n \exp(-(1-\kappa^{(m)})k) u^{T(m)}_l p_l(0)
\label{solution}
\eea
where we have used \req{p} and \req{u}. As $k\to\infty$ only the leading
eigenvalue $\kappa^{(1)}=1$ survives, and determines the asymptotic
value of $p_n(k)$. The next to leading eigenvalue
$\kappa^{(2)}$ will control the rate of convergence. In particular:
\bea
p_n(k)\to p_n(\infty) + \sum_{l\geq2}\exp(-(1-\kappa^{(l)})k)(\Delta p)^{(l)}_n
\label{approach1}
\eea
where we have defined
\beq
(\Delta p)^{(l)}_n =  p^{(l)}_n\sum_m u^{T(l)}_m p_m(0)
\eeq
and
\bea
p_n(\infty) &=&  p^{(1)}_n \sum_l u^{T(1)}_l p_l(0)\none
   &=& p_n^{(1)}
\eea
Thus, the photon distribution converges to its equilibrium value, as
desired.
Moreover, \req{approach1} implies:
\bea
p_n(k)- p_n(\infty) = \sum_{l\geq2}\exp(-(1-\kappa^{(l)})k)(\Delta p)^{(l)}_n
\approx \exp(-(1-\kappa^{(2)})k)(\Delta p)^{(2)}_n + \,.\,.\,.
\eea
and the approach to equilibrium is determined by $\kappa^{(2)}$
if there is no degeneracy in the next to leading eigenvalues.
As we will see in the subsequent numerical analysis, interesting things
occur when $\kappa^{(2)}$ and $\kappa^{(3)}$ are close within the
numerical accuracy of the calculation.

Note that the continuous master equation \req{cont.master} can be integrated
in precisely the same way, with $-(1-S)$ replaced by $-\gamma L$, so that
the asymptotic behaviour is controlled by the eigenvalues $\lambda^{(l)}$
of $-\gamma L$, instead of $1-\kappa^{(l)}$. In the appendix we prove that
these eigenvalues coincide in the large $N$ limit, and this is verified
numerically in the next section.

Before closing this section we make one further comparison between our
analysis and that of \cite{Pers}, who look at correlation functions
of the spin variables of the form
\bea
\gamma_k = \frac{\langle s s \rangle_k - \langle s \rangle ^2}{\langle s^2
\rangle - \langle s\rangle^2}\label{corr}
\eea
where
\bea
\langle s \rangle = \sum_{s=\pm 1} s {\cal P}(s); \,\,\,\,\,\,{\cal P}(s) =
\sum_n S(s)_{nm} p^{(1)}_m = u^{T(1)}_n S(s)_{nm} p^{(1)}_m
\eea
Similarly,  $\langle s s\rangle _k$ is the joint probability for observing the
states of two atoms, $s_1$ followed by $s_2$, with $k$ unobserved atoms between
them:
\bea
\langle s s \rangle_k &=& \sum_{s_1=\pm 1,s_2= \pm 1} s_1 s_2 {\cal
P}_k(s_1.s_2)\nonumber \\
{\cal P}_k(s_1,s_2) &=& u^{T(1)} S(s_2) S^k S(s_1) p^{(1)}\label{ssk}
\eea
Note that the above expectation values have assumed that the photon
distribution is already at its equilibrium value
$p^{(1)}_n$ before the spin of the first atom is measured. In spite of
this, the correlation function \req{corr} does describe the approach
to equilibrium and is directly related to the quantities that we study
in the present paper. In particular,
the correlation length associated with $\gamma_k$ approaches zero
in the limit of large $k$, and this approach
is governed by the same eigenvalues
as the approach to equilibrium of the photon distribution. The basic
argument
is as follows. In effect $\langle ss \rangle_k$ 
depends on the conditional probability that
one first measures the spin to be $s_1$, say and that after $k$ atoms
pass, spin $s_2$ is measured. However, when one measures the first
spin,
one effectively applies
a projection operator which moves the photon distribution
 away from the equilibrium configuration. The shape of this projected
photon distribution then determines the correlation between the first
and second spin measurements.
One expects that for a very large number $k$ of atoms between the two
measurements, the photon distribution again approaches its equilibrium
value, so that correlation between the two spin measurement
vanishes. That is:
\bea
{\rm lim}_{k\rightarrow\infty} \langle s s \rangle _k \rightarrow \langle s
\rangle ^2
\eea
and the correlation function approaches zero as $k$ approaches
infinity.
  The correlations between well separated atoms therefore measure the rate at
which the photon distribution settles
back to equilibrium.

We will now prove the above assertions. We look for exponential decay
of the
correlation function,
\bea
\gamma_k \sim {\rm exp} \left( -\frac{k}{R\zeta}\right)\label{corrlength}
\eea where $k\approx Rt$ and $\zeta$ is the correlation length, or the typical
length of time over which the cavity remembers pumping events.
We can rewrite,
\bea S^k = T (T^{-1} S T)^k T^{-1} = T D^k T^{-1}\label{Sk}
\eea
where
\beq
D_{kl}= \kappa^{k}\delta_{kl}
\label{D}.
\eeq
We want to consider the way in which $D^k$ approaches equilibrium.  We start
from,
\bea
D^{k+1} - D^k = (D-1) D^k
\eea
We consider the limit of large $k$ in order to isolate the exponential
dependence.  We write, $D^{k+1} - D^k \rightarrow d\,D(k)$ and $1=(k+1)-k
\rightarrow d\,k$.  We obtain,
\bea
d\,D_{nm}(k) &=& (D-1)_{nl}D_{lm}(k)\,\,d\,k\nonumber \\
&=& (\kappa^{(n)}-1)\delta_{nl} D_{lm}(k)\,\,d\,k\\
&=& (\kappa^{(n)} -1) D_{nm}(k)\,\,d\,k \label{contD}\nonumber
\eea
which has the solution,
\bea D_{nm}(k) = D_{nm}(0){\rm exp}^{-(1-\kappa^{(n)})k}\label{Dsoln}
\eea
where $D_{nm}(0)$ is an integration constant.  Writing out the first two terms
in the sum over n we obtain,
\bea
D_{nm}(k) = \delta_{n1} D_{nm}(0){\rm exp}^{-(1-\kappa^{(1)})k} +
\Sigma_{n=2}^{\infty} D_{nm}(0){\rm exp}^{-(1-\kappa^{(n)})k}
\eea
Since $\kappa^{(1)}=1$ and $D$ is diagonal we have,
\bea
D_{nm}(k) = \delta_{n1}\delta_{m1} D_{nm}(0) + \Sigma_{n=2}^{\infty}
D_{nm}(0){\rm exp}^{-(1-\kappa^{(n)})k}\label{Dk}
\eea
Substituting \req{Sk} and \req{Dk} into \req{ssk}
we have,
\bea
\langle s s \rangle_k &=& \Sigma_{s_1,s_2} s_1 s_2  u_n^{T(1)} S_{nj}(s_2)
T_{jq}\nonumber \\
&&\left[ \delta_{q1}\delta_{m1} D_{qm}(0) + \Sigma_{q=2}^{\infty} D_{qm}(0){\rm
exp}^{-(1-\kappa^{(q)})k}\right]\label{SB}
T_{mr}^{-1} S_{rl}(s_1) p_l^{(1)}\nonumber
\eea
Denoting the first (leading order) term in the square
brackets by $\langle s s \rangle_k ^{l.o.}$, we find
\bea
\langle s s \rangle_k ^{l.o.}&=& \Sigma_{s_1,s_2} s_1 s_2  u_n^{T(1)}
S_{nj}(s_2) T_{j1}D_{11}(0)T_{1r}^{-1} S_{rl}(s_1)p_l^{(1)}\nonumber \\
&=&\Sigma_{s_1,s_2} s_1 s_2  u_n^{T(1)} S_{nj}(s_2) p_j^{(1)} D_{11}(0)
u_r^{T(1)}S_{rl}(s_1)p_l^{(1)}
\eea
where we have used \req{p} and \req{u}.  Since $\kappa^{(1)}=1$ we have from
\req{Dsoln} that $D_{11}$ is independent of $k$ and from \req{D} that $D_{11}
=1$.
This gives,
\bea
\langle s s \rangle_k ^{l.o.}= \Sigma_{s_1,s_2} s_1 s_2  (u_n^{T(1)}
S_{nj}(s_2) p_j^{(1)})(u_r^{T(1)}S_{rl}(s_1)p_l^{(1)}) = \langle s \rangle ^2
\eea
The second term in square brackets of \req{SB} shows that, as stated earlier,
all of the eigenvalues other than $\kappa^{(1)}$ must be less than one, or the
correlation length diverges.
Recalling that the eigenvalues are labelled by size, the leading order non-zero
term in the correlation function \req{corr} has the form,
\bea \gamma_k \sim e^{-( 1-\kappa^{(2)})k}
\label{approach2}
\eea
and from \req{corrlength} we have,
\bea
R\zeta \sim \frac{1}{ (1-\kappa^{(2)})}
\label{correlation 2}
\eea
Comparing \req{approach1} and \req{approach2} we see that the
correlation length $\zeta$ is determined by the same eigenvalue that
determines the approach to equilibrium, as claimed.

\section{ Numerical Results}

We wish to investigate numerically the approach to equilibrium of the
photon distribution as described by the dynamical master equation
\req{master2}, with $S$ given by \req{S}. We assume for concreteness
that before the first atom enters the cavity, the photons are in
thermal equilibrium at the temperature $T$ which characterizes the
thermal properties of the system throughout the experiment. In
principle the asymptotic properties of the approach to equilibrium
should not be sensitive to the initial photon distribution, but we
observed some interesting short time behaviour which presumably does
depend on the initial distribution.  The short time behaviour may thus
have physical relevance. To begin, we start with the photon
distribution: 
\bea p_n = [1-e^{-{\hbar \omega\over kT}}]e^{-{n \hbar
    \omega\over kT}}
\label{BE}
\eea 
The mean photon number is therefore \bea n_b = \Sigma_n n p_n =
\frac{1}{e^{\hbar \omega\over kT} -1} \eea Note that the mean photon
number also plays an important role in the master equation, i.e. in
the matrix $L_C$ (\req{LC}) that determines the rate at which the
photon distribution relaxes into thermal equilibrium.  For a typical
two level Rydberg atom \cite{Rempe}, $\hbar\omega\approx 1.4\times
10^{-23} J$, so that 
\bea T (\hbox{Kelvin}) \approx
{1\over\ln{1+1/n_b}} 
\eea 
Typical experimental temperatures range
between $T=0.4K$ ($n_b=0.1$) and $T=10K$ ($n_b=10$). In terms of
$n_b$, the thermal distribution is: 
\bea p_n =
\frac{1}{1+n_b}\left(\frac{n_b}{1+n_b}\right)^n
\label{thermal}
\eea

As shown in Section 3 above, the asymptotic behaviour of the approach
to equilibrium and the long time correlation functions are both
determined by the leading eigenvalues of the matrix $S$. Fig. 1 shows
how the correlation length \req{correlation 2} changes, for
$n_b=3$, with pumping
rate and interaction time
\footnote{In order to calculate the eigenvalues
we truncated the photon number at $n=200$, so that $S$ was
a $200\times200$ matrix.}. The axes correspond to the pumping rate $N$
and $\theta = g\tau\sqrt{N}$, which is a scaled time parameter that is
useful in revealing the phase structure.  One can readily see the
evidence for critical points at $\theta\sim1$ and $\theta\sim 4.6$,
which mark the transition from the thermal phase to the maser phase
and from the thermal phase to the first critical phase, respectively.
As anticipated the phase structure matches the one obtained by
\cite{Pers} using the continuous master equation.
\par
In the present work, we implement \req{master2} directly by doing the
numerical ``experiment'' of sending in one atom after another (i.e.
multiplying $p$ by $S$), and seeing how the photon distribution
changes with $k$, the number of atoms that have passed through the
cavity.  The purpose of the numerical experiment was to measure how
long it takes to get to equilibrium for different values of the
physical parameters $N$ and $n_b$\footnote{In the subsequent analysis
  $g\tau$ is kept fixed, which is relevant experimentally. The
  effect of varying $g\tau$ will be examined in future work.}.  This
could be important in physical experiments in which the results are
interpreted in terms of the equilibrium photon distribution. In
general we found that convergence was very rapid, with some
interesting anomalies in the short time behaviour \footnote{Short
  ``time'' here actually refers to the first few atoms in the
  iteration.}.
\par
In order to deal with finite dimensional matrices, we need to truncate
the photon distribution at $n=n_{max}$, say. Consistency requires that
the probability of having $n_{max}$ photons in the cavity be small
compared to the numerical accuracy of the calculation. This was
checked by calculating the normalization of the probability
distribution after each iteration. We found that the slight error in
the normalization grew geometrically with the number of iterations,
which was problematic for runs that contained thousands of atoms. We
found however that this behaviour could be corrected by simply
re-normalizing $p(k)$ at each iteration. If this was done, the errors
grew only linearly with $k$, and $n_{max}$ of about 200 was
sufficiently large for our purposes.

The purpose of the numerical experiment was to measure how long it
takes to reach equilibrium for different values of the physical
parameters $g\tau$, $N$ and $n_b$. This could be important in physical
experiments in which the results are interpreted in terms of the
equilibrium photon distribution. In general we found that convergence
was very rapid.  In order to have a quantitative measure of ``how
close'' the system is to equilibrium, it is necessary to define a
suitable measure on the space of photon distributions, which for the
present purposes could be thought of as an $n_{max}$ dimensional vector
space. We therefore define the distance between the photon
distribution after $k$ atoms and the equilibrium distribution by: 
\bea
\Delta p_n(k) = |p_n(k) - p_n^{equilib}|
\eea
and take as a test for convergence the condition that
all $p_n(k)$ must be within a certain range of the equilibrium
value: 
\bea
\max \Delta p_n(k) < 0.005
\eea
In order to have a point of comparison, we
also checked convergence with a different measure, namely: 
\bea
\Delta {\cal P}_+(k) = |{\cal P}_+(k) - {\cal P}_+(equil)|
\eea
where ${\cal P}_+(k)$ and ${\cal P}(equil)$ are the probability for an
atom entering emerging from the cavity in the excited state
for photon distributions during the transit given by $p_n(k)$ and
$p_n^{eq}$, respectively. (cf. \req{eq: P+}.)
We then used as a second test for convergence
\bea
\Delta {\cal P}_+(k) < 0.005
\eea
This test 
compares the probability that the $kth$ atom will emerge in the
excited state to the same probability at equilibrium. It therefore has
direct physical relevance.  Figs. 2a) and 2b) plot $\Delta p(k)$ and
$\Delta P_+(k)$ as functions of $k$ for $N=45$, $n_b=3.0$ and
$\tau=1.0$. It is interesting that the system first moves towards
equilibrium (very rapidly for $\Delta {\cal P}_+(k)$) and then moves
away from equilibrium before it settles into its exponential approach
to equilibrium. This feature appears to be fairly generic for large
$N$.
\par

In order to do a systematic analysis of the convergence rate as a
function of the physical parameters, we define $k_{max}$ as the number
of atoms it takes for $\Delta p(k)$ to get to some critical value,
$\Delta_{crit}$, or less.  For small enough $\Delta_{crit}$, $k_{max}$
should be large, in which case the convergence will be determined by
the next to leading eigenvalue of $S$.  Fig. 3 plots $k_{max}$ as
obtained from the two tests as functions of $\theta$ and $N$ for
$n_b=3$, with the condition $\Delta_{crit}= 0.005$.  Clearly the
resulting values of $k_{max}$ are large enough to be in the region of
asymptotic convergence and the phase structure is the same as that
predicted analytically using the eigenvalues of $S$. This result
confirms the validity of our numerical method.
\par
We now use the numerical experiments to investigate a different but
related property of convergence. In particular, we look at how
$k_{max}$ is affected by a change in $n_b$ and $N$ for fixed
interaction time. As shown in Fig. 4, as $n_b$ gets large, the
convergence is uniformly rapid for all $N$, whereas for low $n_b$
(i.e. low temperatures) there are critical values of $N$ for which
convergence slows suddenly. 
These are presumably the same transitions as
in Fig. 1 but seen from a different view. In particular, lines of fixed $n_b$
and $g\tau$ correspond to a section of Fig. 1 with 
$\theta = g\tau\sqrt{N}.$ Fig. 5 shows this structure for fixed
$n_b$ and $g\tau$. As shown in Fig. 6, these ``steps'' in the
convergence coincide (at least approximately) with values of the
parameters in which the correlation length, as determined by the
leading eigenvalue, increases.  Fig.~6 plots the correlation lengths
for $n_b=1$ for the first four leading eigenvalues. It is interesting
that the ``steps'' correspond to points where the eigenvalues appear
to cross. However, the eigenvalues are not degenerate \cite{Pers},
so the curves do not actually intersect. In
Fig. 7 the equilibrium photon distributions are plotted for $nb=1$ as
a function of $N$. It shows clearly that the ``crossing'' of the
eigenvalues is related to a discrete shift in the peak of the
equilibrium photon distribution. The crossings, and the associated
transitions in the convergence rate illustrated in Fig.~5, occur when
the photon distribution is in the process of shifting, i.e. where
there are two peaks.

\section{Conclusions}

We have presented a systematic analysis
of the discrete master equation describing the approach to equilibrium
of the  micromaser in the presence of thermal dissipation. As expected,
the long time behaviour is determined by the leading eigenvalues of
the discrete transformation matrix. Interestingly, the eigenvalues of the
matrix, evaluated numerically, become at some places
nearly degenerate, and the phase structure
of the micromaser occurs at or near where the eigenvalues
come close to crossing.  Our
analytic results
 confirm general features that emerge from the continuous master
equation in \cite{Pers}. We also have examined the
approach to equilibrium of the
micromaser both at short and long times using numerically methods.
Our numerical results are consistent with the
behaviour expected from the leading eigenvalues of both the discrete
and continuous transformation matrices. However, our results also
show some interesting feature for short times; in particular,
the system in general first approaches
equilibrium relatively rapidly, then moves
away from equilibrium, and then finally
settles into its exponential approach
to equilibrium. This behaviour appears to be fairly
generic for large values of $N$, but 
more analysis is required to determine the source and relevance of these
short time features. Moreover we have, in the numerical experiments, kept
the transit time fixed. In future work we hope to examine how varying
$g\tau$ affects the above mentioned features.\\[5pt]
\par\noindent
{\bf Acknowledgements}
\par
 This work
was supported in part by the Natural Sciences and Engineering
Research
Council of Canada, and by Career Focus, Manitoba.

  \par\vspace*{20pt}
{\bf Appendix: The Relationship Between the Continuous and Discrete Cases}

We expect differences in the dynamical behaviour in the discrete and
continuous formalisms, but there should exist a limit in which the two
formalisms coincide.  We look at the discrete formalism in the large flux
limit.  We take $k = Rt$ to be large, which means that $t\gg 1/T$, or that the
total time over which the system is observed is much greater than the time
between individual atoms.

To take the large flux limit of the discrete master equation, we follow the
derivation of \req{contD}. The discrete master equation has the form \req{S},
\bea
p^{k+1} = S p^k
\eea
We write,
\bea
p^{k+1} - p^k &=& (S-1)p^k\nonumber \\
  (p^{k+1} - p^k)            &=& T[T^{-1}(S-1)T] T^{-1} p^k\nonumber \\
p_b^{k+1} - p_b^k &=& T_{bn}[(\kappa^{(n)} -1)\delta_{nm}](T^{-1} p^k)_{ma}
\eea
which can be written in differential form as,
\bea
d\,p = -(1-S) p d\,k\label{difd}
\eea
Since $k=Rt$ we can write $d\,k = Rd\,t$ and compare \req{L} and \req{difd}:
\bea
-\gamma L &=& -(1-S) R \nonumber \\
L &=& N(1-S);\,\,\,\,\,\,R/\gamma = N
\eea
Using \req{S} and \req{L} we find,
\bea
L_C-N(M-1) = {\tilde N}-{\tilde N}\frac{1}{1+L_C/{\tilde N}} M
\eea
We study the behaviour of the next to leading eigenvectors, since the
corresponding eigenvalues control the behaviour of the approach to equilibrium.
We have,
\bea (L_C-N(M-1))p^{(2)} = \lambda^{(2)} p^{(2)}\label{1}
\eea
and \bea
S\tilde{p}^{(2)} &=&\kappa ^{(2)} \tilde{p}^{(2)}\nonumber \\
\rightarrow \frac{1}{1+L_C/\tilde{N}} M \tilde{p}^{(2)} &=& \kappa^{(2)}
\tilde{p}^{(2)}\nonumber \\
\rightarrow (L_C - \frac{\tilde{N}(m-1)} {\kappa^{(2)}}) \tilde{p}^{(2)} &=&
\tilde{N}[\frac{1}{\kappa^{(2)}} -1 ]\tilde{p}^{(2)} \label{2}
\eea
Comparing \req{1} and \req{2} we have,
\bea
N=\tilde{N}/\kappa^{(2)};\,\,\,\,\,\,\,\,\,\,\,\lambda^{(2)} =
\tilde{N}[\frac{1}{\kappa^{(2)}}-1]
\eea
Solving this set of equations we have,
\bea
\kappa^{(2)} = 1-\frac{\lambda^{(2)}}{N}\nonumber \\
{\rm ln} \kappa^{(2)} = {\rm ln}(1-\frac{\lambda^{(2)}}{N}) \sim
-\frac{\lambda^{(2)}}{N} \nonumber \\
-R{\rm ln}\kappa^{(2)} \sim \frac{R}{N} \lambda^{(2)} = \gamma \lambda^{(2)}
\eea
where we have taken $N$ large, or $1/\gamma \gg T$, which means that the
typical photon decay time is much greater than the typical time between atoms.

\clearpage
\section*{Figures}
\vspace*{15pt}
\par
\begin{figure}[hb]
\leavevmode
\epsfxsize=6.5 in
\epsfbox{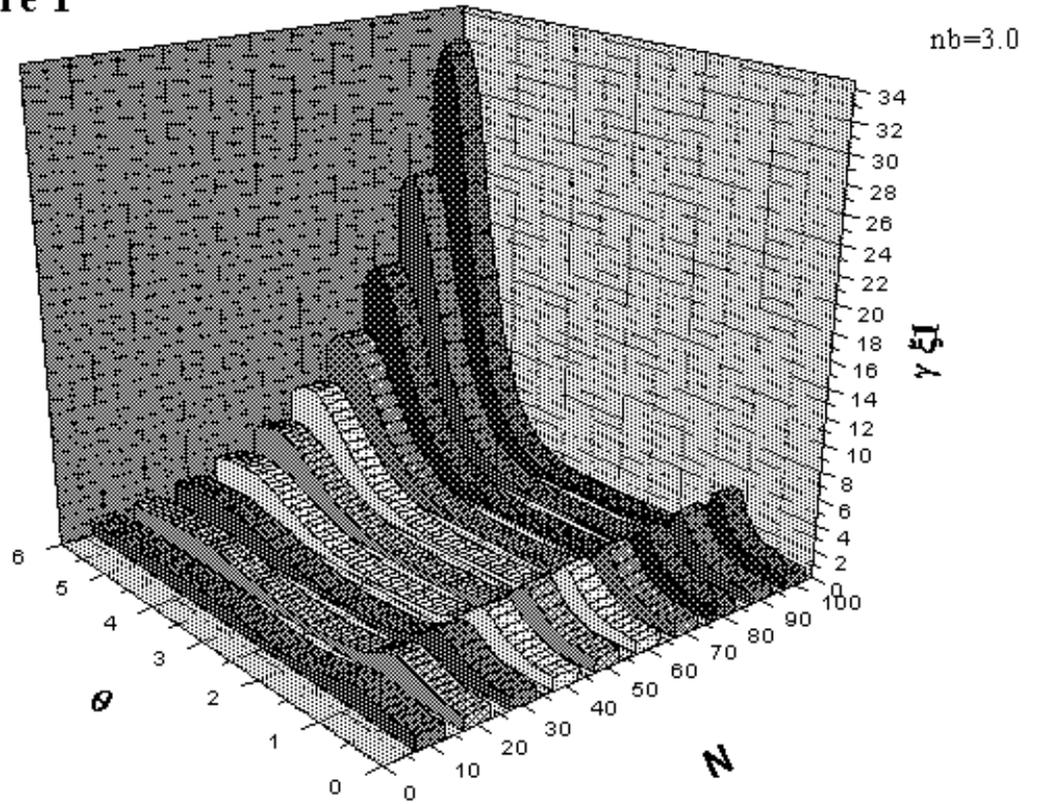}
\caption{Variation of ccorrelation length with 
pumping rate and interaction time for fixed $n_b=3$}
\end{figure}
\par
\begin{figure}[hb]
\leavevmode
\epsfxsize=6.5 in
\epsfbox{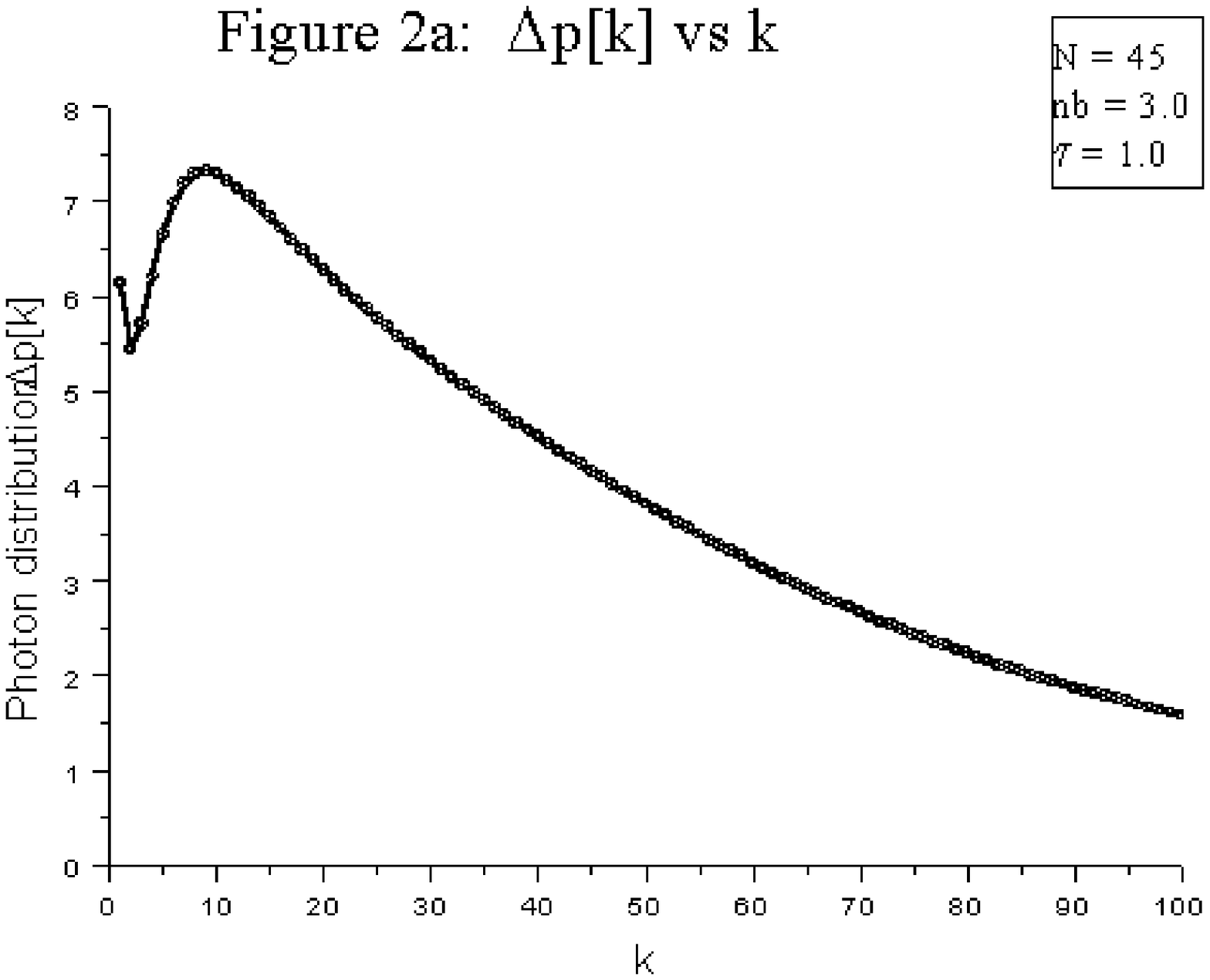}
\epsfxsize=6.5 in
\epsfbox{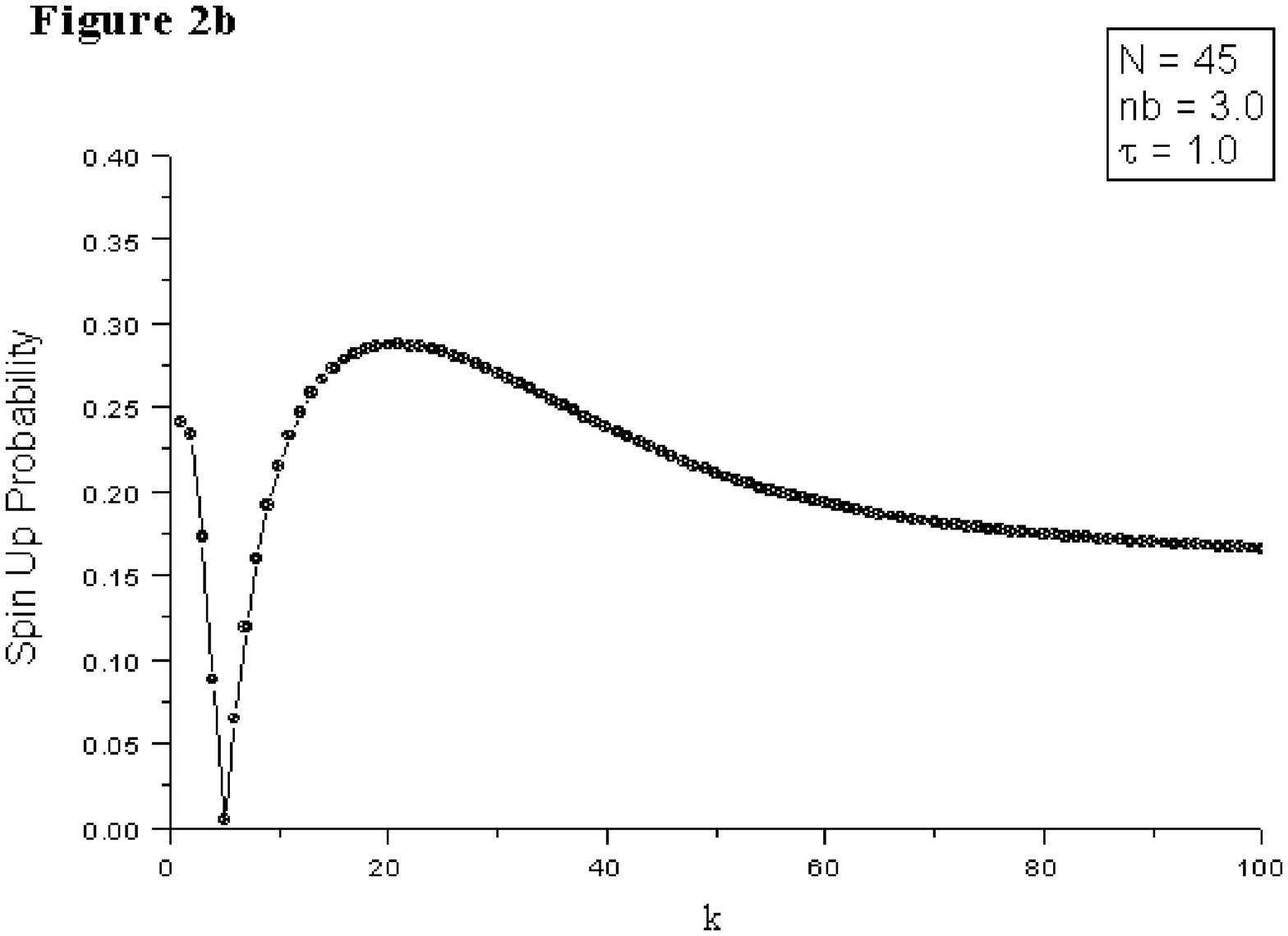}
\caption{Variation of (a) $\Delta p(k)$ and (b) $\Delta P_+(k)$ with $k$}
\end{figure}
\par
\par
\begin{figure}[hb]
\leavevmode
\epsfxsize=6.5 in
\epsfbox{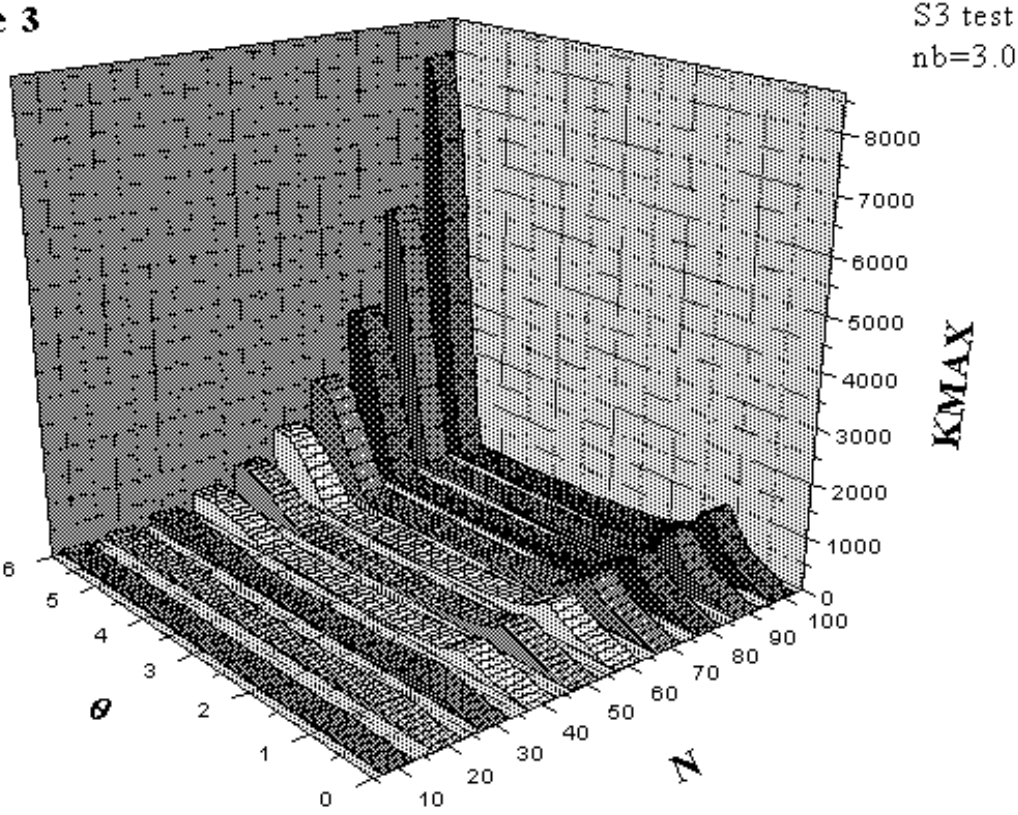}
\caption{Variation of $k_{max}$ with $\theta$ and $N$ for
fixed $n_b=3$ for the two tests described in the text.}
\end{figure}
\par
\par
\begin{figure}[hb]
\leavevmode
\epsfxsize=6.5 in
\epsfbox{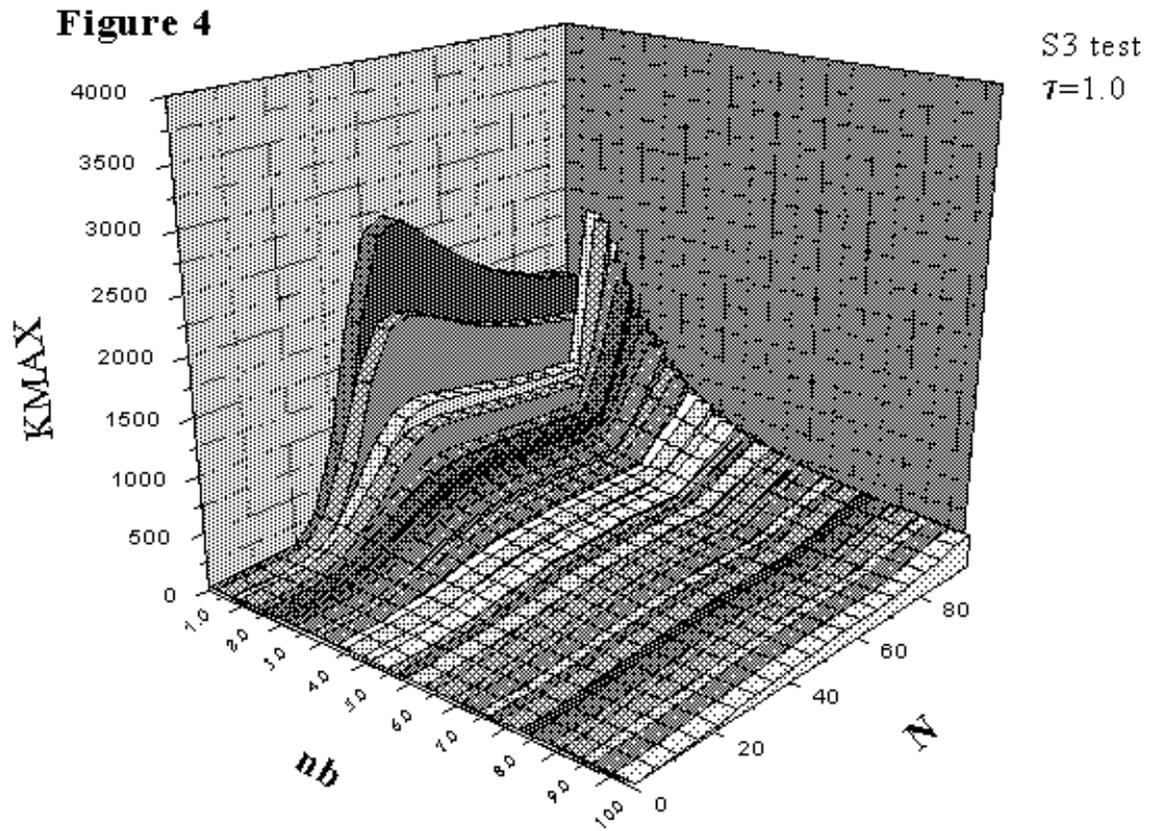}
\caption{Variation of $k_{max}$ with $n_b$ and $N$ for
fixed interaction time.}
\end{figure}
\par
\par
\begin{figure}[hb]
\leavevmode
\epsfxsize=6.5 in
\epsfbox{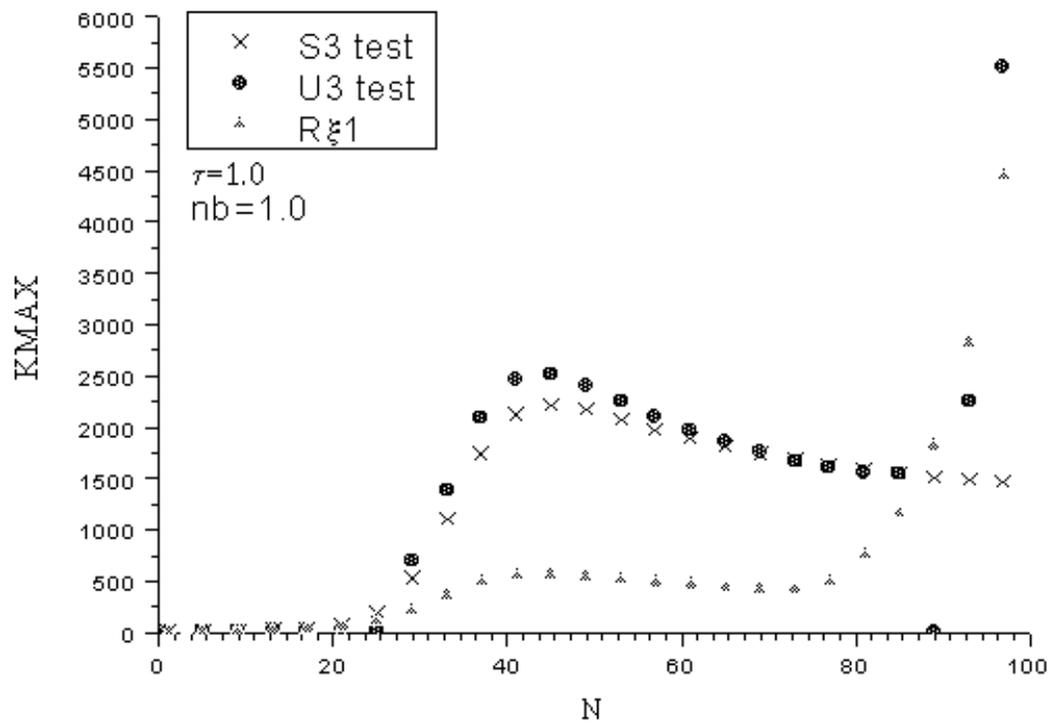}
\caption{Variation of $k_{max}$ with $N$ for
fixed $n_b$ and $g\tau$.}
\end{figure}
\par
\par
\begin{figure}[hb]
\leavevmode
\epsfxsize=6.5 in
\epsfbox{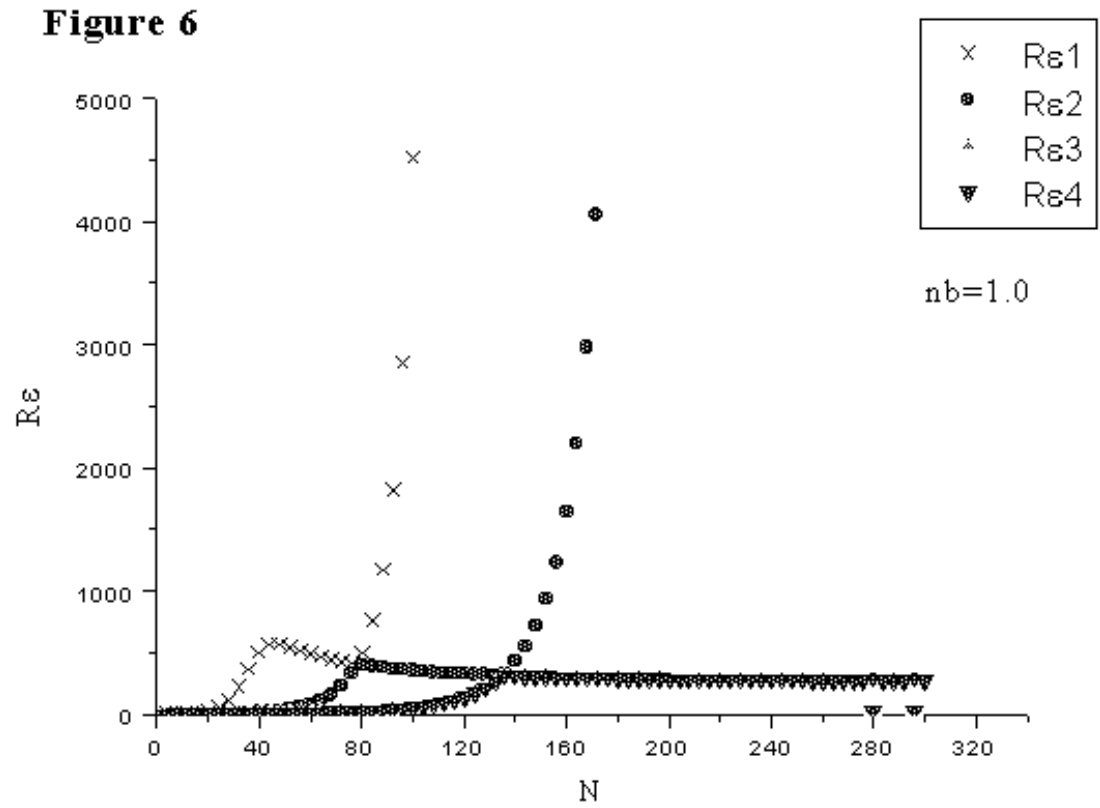}
\caption{Variation of correlation lengths with $N$ for $n_b=1$ 
for the first four leading eignevalues.}
\end{figure}
\par
\par
\begin{figure}[hb]
\leavevmode
\epsfxsize=6.5 in
\epsfbox{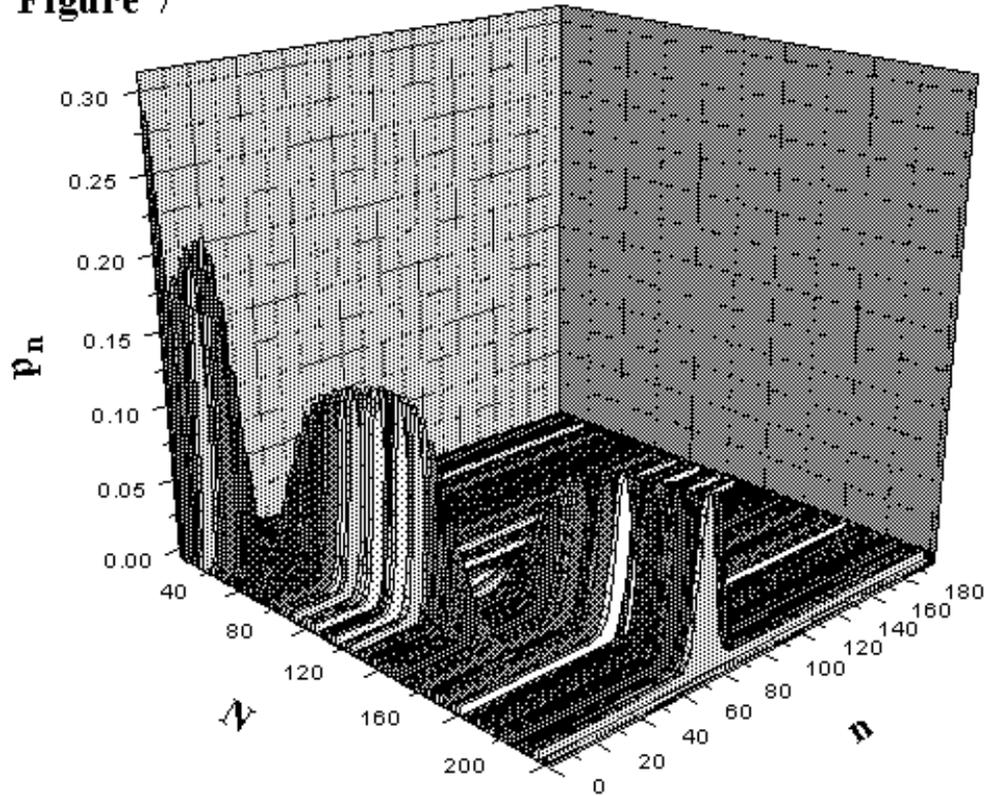}
\caption{Plots of the equilibrium photon distributions for $n_b=1$
as a function of $N$}
\end{figure}
\par

\end{document}